\documentstyle[aps,epsfig,multicol]{revtex}

\newcommand{\simge}
{\raisebox{-0.75ex}[-1.5ex]{$\;\stackrel{>}{\sim}\;$}}

\def\s{{\sigma}}
\def\e{{\epsilon}}
\def\k{{ {\bf k} }}

\def\q{{ {\bf q} }}
\def\Q{{ {\bf Q} }}
\def\B{{ {\bf B} }}

\def\w{{\omega}}
\def\a{{\alpha}}

\begin{document}

\def\runtitle{
Magnetic-Field-Induced Antiferromagnetism 
in Two-Dimensional Hubbard Model
}
\def\runauthor
{Keitaro {\sc Sakurazawa},
Hiroshi {\sc Kontani} and
Tetsuro {\sc Saso}}

\title{
Magnetic-Field-Induced Antiferromagnetism \\
in Two-Dimensional Hubbard Model: 
Analysis of CeRhIn$_5$
}

\author{
Keitaro {\sc Sakurazawa}$^1$,
Hiroshi {\sc Kontani}$^2$ and
Tetsuro {\sc Saso}$^1$
}

\address{
$^1$Department of Physics, Saitama University,
255 Shimo-Okubo, Saitama 338-8570, Japan.\\
$^2$Department of Physics, Nagoya University,
Furo-cho, Nagoya 464-8602, Japan.
}

\date{\today}

\maketitle

\begin{abstract}
We propose the mechanism for the magnetic-field-induced 
antiferromagnetic (AFM) state in a two-dimensional 
Hubbard model in the vicinity of the AFM quantum
critical point (QCP), using the fluctuation-exchange 
(FLEX) approximation by taking the Zeeman energy due to 
the magnetic field ${\bf B}$ into account.
In the vicinity of the QCP,
we find that the AFM correlation 
perpendicular to ${\bf B}$ is enhanced, 
whereas that parallel to ${\bf B}$ is reduced.
This fact means that the finite magnetic field
increases $T_{\rm N}$, with the AFM order
perpendicular to ${\bf B}$.
The increment in $T_{\rm N}$ 
can be understood in terms of the
reduction of both quantum and thermal 
fluctuations due to the magnetic field,
which is caused by the self-energy effect
within the FLEX approximation.
The present study naturally explains
the increment in $T_{\rm N}$ in CeRhIn$_5$ 
under the magnetic field found recently.
\end{abstract}

\noindent
{KEY WORDS: field-induced magnetism,
FLEX, antiferromagnetic fluctuations,
CeRhIn$_5$}

\sloppy

\begin{multicols}{2}

Recently, critical phenomena in the vicinity of the magnetic 
quantum critical point (QCP) have attracted much interest 
in strongly correlated metals.
Experimentally, the outer magnetic field is frequently 
used to change the distance from the QCP.
As for the antiferromagnetic (AFM) QCP,
the magnetic field is believed to increase the 
distance to the QCP in general.
Spin fluctuation theories
such as the SCR theory
 \cite{Moriya}
and the fluctuation-exchange (FLEX) approximation
 \cite{Bickers},
have succeeded in describing various
critical phenomena in metals close to the AFM-QCP,
such as the non-Fermi liquid-like 
behaviors of various transport coefficients
 \cite{Kontani-Hall,Kontani-Nernst}.
However, 
previous studies on the effect of the magnetic field
based on the spin fluctuation theory
were not comprehensive
 \cite{Shioda,Miyake}.

CeMIn$_5$ (M=Rh, Co, or Ir)
is a well-known quasi-two-dimensional
heavy fermion compound, where single conductive
CeIn layers stack perpendicular to the $c$-axis.
CeCoIn$_5$ is 
a superconductor with $T_{\rm c}=2.3$ K
at ambient pressure
 \cite{CeCoIn}.
In CeRhIn$_5$,
the AFM order emerges at $T_{\rm N}=3.8$ K 
at ambient pressure,
and the superconductivity emerges
at $T_{\rm c}\approx 2$K below $P=1.6$ GPa
 \cite{TN1,TN2}.
Recent experiments reveals that
the $T_{\rm N}$ {\it increases}
under the magnetic field along the $a(b)$-axis.
When $B=9$ T,
the increment in $T_{\rm N}$ is approximately 0.15 K.
A small increment in $T_{\rm N}$ is
also observed in Ce$_2$RhIn$_8$
which is composed of double CeIn layers.
However, there has been no 
theoretical explanation for this phenomenon.

In the present study,
we investigate the two-dimensional Hubbard model
under the uniform magnetic field $\B$ along the $x$-axis,
based on the FLEX approximation.
In the vicinity of the AFM-QCP,
we find that the AFM spin correlation of 
the $y(z)$-component is {\it enhanced} by the applied magnetic field.
In the obtained phase diagram,
the magnetic transition temperature $T_{\rm N}$,
below which the staggered magnetism emerges on the $yz$-plane,
increases with magnetic field.
The mechanism of the field-induced antiferromagnetism (FI-AFM) 
proposed in the present study
will be universal in low-dimensional metals close to the AFM-QCP,
contrary to the fact that
the magnetic field suppresses $T_{\rm N}$ in usual models
by the mean-field approximation.
The present study naturally explains
the enhancement in $T_{\rm N}$ under the magnetic field
in CeRhIn$_5$.

\begin{figure}
\begin{center}
\epsfig{file=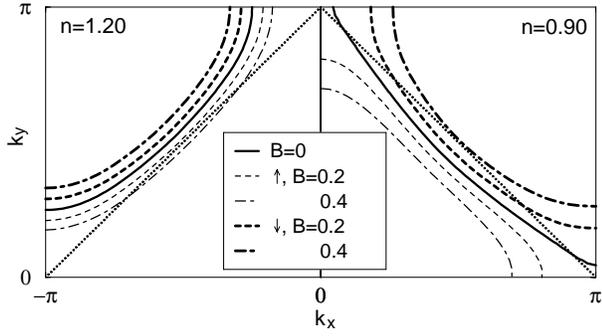,width=8cm}
\end{center}
\caption{
FS's for $\uparrow$- and $\downarrow$-electrons
under various magnetic fields for $n=0.90$ and $1.20$
at $T=0.02$.
}
  \label{fig:FS}
\end{figure}

We analyze the following two-dimensional Hubbard model:
\begin{eqnarray}
H= \sum_{\k\s}\e_{\k\s}c_{\k\s}^\dagger c_{\k\s}
 + U\sum_{\k\k'\q}c_{\k+\q\uparrow}^\dagger
 c_{\k'-\q\downarrow}^\dagger c_{\k'\downarrow} c_{\k\uparrow},
\end{eqnarray}
where $\s=1(-1)$ corresponds to the $\uparrow$- ($\downarrow$-)
spin state and $\e_{\k\s}= \e_\k +\s B$,
where the factor $\s B$ represents the Zeeman energy.
The spin quantization axis is the $x$-axis.
We study the square lattice tight-binding model 
with nearest neighbor hopping ($t$)
and next-nearest neighbor hopping ($t'$).
The dispersion of the electron is given by
$\e_\k= -2t(\cos k_x + \cos k_y ) - 4t'\cos k_x \cos k_y$.
We study the case of 
$(t,t')=(1,-0.25)$ with the electron density
$n=0.90$ ($n=1.20$) per site, which corresponds to a 
hole-doped (electron-doped) high-$T_{\rm c}$ cuprates.
In the case of $n=0.90$,
the Fermi surface(FS) is very close to 
the van-Hove singular point (at $(\pi,0)$ in this case;
see Fig. \ref{fig:FS}), and it is similar to 
the largest (main) cylindrical FS in CeMIn$_5$ (M=Co,Ir,Rh)
 \cite{Maehira}.
Assuming a similar single cylindrical FS,
many aspects of CeMIn$_5$, particularly the 
$d_{\rm x^2\mbox{-}y^2}$-wave superconductivity,
can be reproduced by the perturbation study
 \cite{Nisikawa,Yamada-rev}.

In the presence of the magnetic field along the $x$-axis,
the dynamical spin susceptibilities
within the FLEX approximation
(or random-phase approximation (RPA)),
$\chi_{x}^s(q)$ and $\chi_{y(z)}^s(q)$,
are given by
\begin{eqnarray}
& &\chi_{y}(q)=\chi_{z}(q)
 = \left( \chi_{\uparrow,\downarrow}(q)
 +\chi_{\downarrow,\uparrow}(q) \right)/4
 \label{eqn:chi-xx}\\
& & \chi_{x}(q)= [\chi_{\uparrow,\uparrow}(q)
                   +\chi_{\downarrow,\downarrow}(q)]/4
    +U\chi_{\uparrow,\uparrow}(q)\chi_{\downarrow,\downarrow}^0(q)/2,
 \\
& &\ \ \chi_{\s,-\s}(q)
 = \frac{\chi_{\s,-\s}^0(q)}
        {1-U\chi_{\s,-\s}^0(q)},
 \\
& & \ \ \chi_{\s,\s}(q)
 = \frac{\chi_{\s,\s}^0(q)}
        {1-U^2\chi_{\s,\s}^0(q)\chi_{-\s,-\s}^0(q)},
 \\
& & \ \ \chi_{\s,\s'}^0(q)
 = -T\sum_k G_\s(k+q)G_{\s'}(k).
 \label{eqn:chi0}
\end{eqnarray}
Note that $\chi_{\uparrow,\downarrow}(q)= 
\{\chi_{\downarrow,\uparrow}(q)\}^\ast$.
Here and hereafter, we promise that
$q\equiv(\q,i\w_n)=(\q,2\pi i nT)$ and
$k\equiv(\k,i\e_n)=(\k,\pi i (2n+1)T)$.
Apparently,
both $\chi_{x}(q)$ and $\chi_{y(z)}(q)$
are even functions of $B$,
reflecting the reflectional symmetry in spin space.
Apparently, $\chi_{x}(q)= \chi_{y}(q)$
when $B=0$.

The self-energy in the FLEX approximation 
is given by
\begin{eqnarray}
\Sigma_{\s}(k)
&=&
 U^2 T\sum_{q}[G_{\s}(k-q) (\chi_{-\s,-\s}(q)
                     -\chi_{-\s,-\s}^0(q))
 \nonumber \\
& &+ G_{-\s}(k-q) \chi_{\s,-\s}(q)]
 +Un_{-\s}
 \label{eqn:Sigma},
\end{eqnarray}
where $n_{\s}= T\sum_{k}{\rm Im}G_{\s}(k)
e^{-i\e_n\cdot 0^+}/\pi$
is the density of electrons with $\s$-spin.
Here, we solve the Eqs.
(\ref{eqn:chi-xx})-(\ref{eqn:Sigma})
together with the Dyson equation
$G_{\s}^{-1}(k)= i\e_n+\mu-\e_\k-\s B -\Sigma_\s(k)$
numerically, by adjusting the chemical potential $\mu$
so that $n = \sum_{\s}n_{\s}$.

\begin{figure}
\begin{center}
\epsfig{file=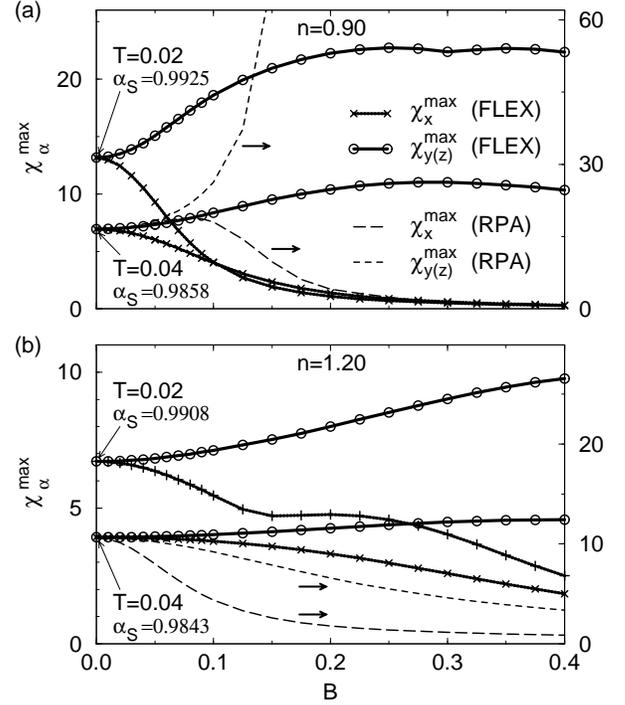,width=8cm}
\end{center}
\caption{
Field dependences of $\chi_{x}^{\rm max}$ and $\chi_{y}^{\rm max}$
by the FLEX approximation (RPA),
under conditions (a) $n=0.90$, $U=5$ ($U=2.10$) and 
(b) $n=1.20$, $U=8$ ($U=2.96$), at $T=0.02$.
In the numerical study by RPA, 
$256\times256$ $\k$-meshes and 256 Matsubara frequencies 
are used.
}
  \label{fig:chi-FLEX}
\end{figure}

\begin{figure}
\begin{center}
\epsfig{file=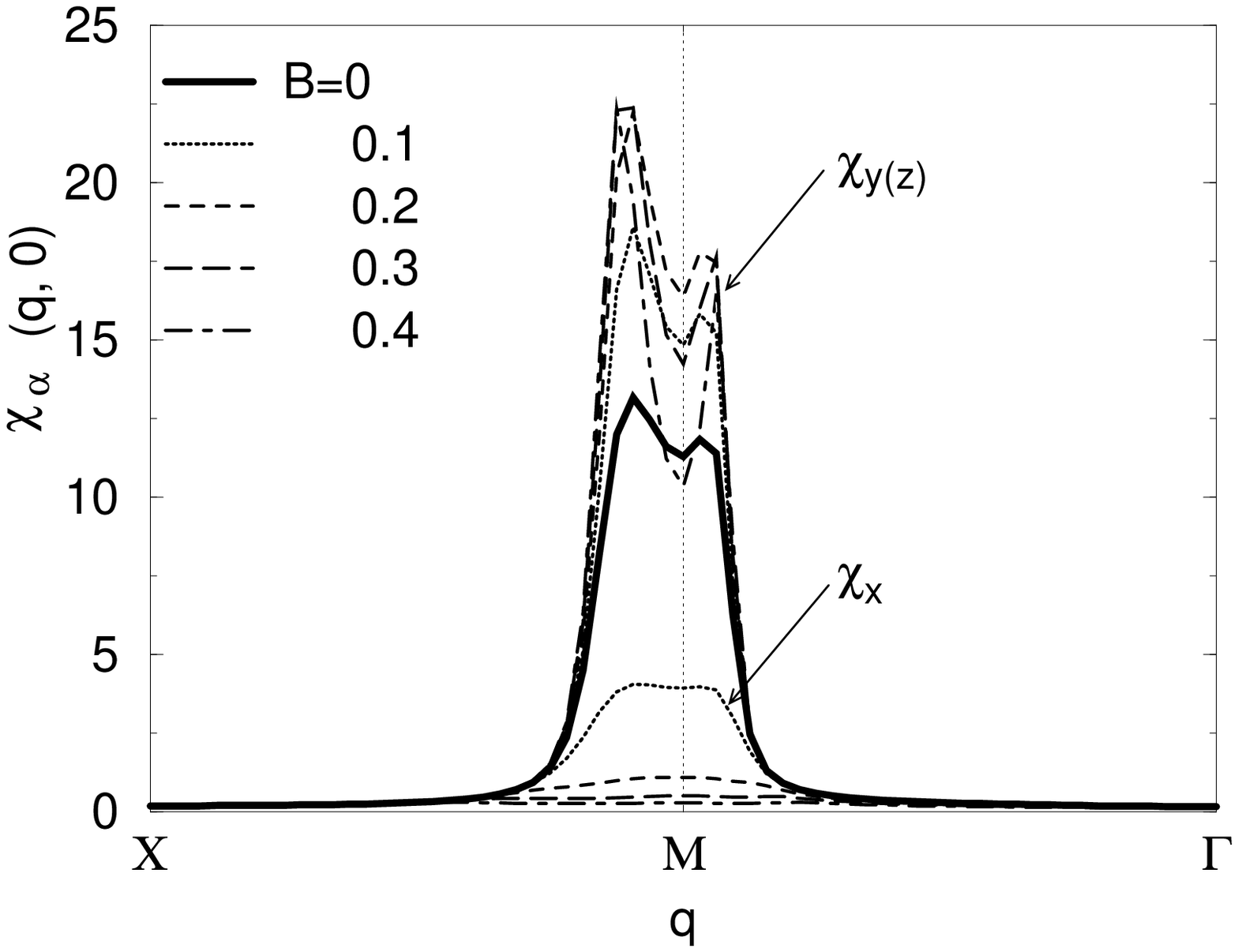,width=8cm}
\end{center}
\caption{
$\chi_{x}(\q,0)$ and $\chi_{y}(\q,0)$ 
under finite $B$ for $n=0.90$ at $T=0.02$.
}
  \label{fig:Q}
\end{figure}

Here, we discuss the numerical results
obtained by the FLEX approximation.
We use $64\times64$ $\k$-meshes and 1028 Matsubara
frequencies in the present numerical study by FLEX approximation.
Figure \ref{fig:chi-FLEX} shows the obtained
static staggered spin susceptibilities:
$\chi_{\a}^{\rm max} \equiv \max_{\q} \chi_{\a}(\q,0)$,
where $\a=x,y,z$.
$\a_{\rm S}\equiv \max_\q U\chi^0(\q,0)$
is the Stoner factor without $B$.
In the FLEX approximation,
$\a_{\rm S}<1$ is always satisfied at finite $T$
in two-dimensional systems,
so the Marmin-Wagner-Hohenberg theorem is satisfied
 \cite{Kino-Kontani,Kino-Kontani2}.
The momentum dependence of $\chi_{\a}(\q,0)$
($\a=x,z$) and the splitting of the FS
under the magnetic field are given in
figs.\ref{fig:Q} and \ref{fig:FS}, respectively,
in the case of $n=0.90$.

In Fig. \ref{fig:chi-FLEX},
$\chi_{x}^{\rm max}$ decreases 
whereas $\chi_{y}^{\rm max}$ {\it increases} 
with ${\bf B}\parallel {\hat x}$
in both cases of $n=0.90$ and $n=1.20$
by FLEX approximation.
Their field dependence becomes more prominent
as $U$ increases, that is, as $\a_{\rm S}$ 
approaches unity.
These results indicate that 
the distance to the AFM-QCP decreases
owing to the uniform magnetic field.
In the FLEX approximation,
the field dependence of the susceptibility
is caused by (i) the change in the nesting conditions
owing to the Zeeman splitting of the FS,
and (ii) the self-energy effect (or mode-mode coupling effect)
which represents the reduction in $\chi^{\rm max}$ 
and its Curie-Weiss-like temperature dependence
owing to the spin-fluctuations.
In the FLEX approximation,
a large Im$\Sigma(\k,-i\delta)$ caused by spin fluctuations
reduces the density of states (DOS) at $\mu$,
which makes $\chi^{\rm FLEX}\gg\chi^{\rm RPA}$.
Below, we will discuss that the effect (ii),
which is absent in the RPA
is important to explain  
why $\chi_{y(z)}^{\rm max}$ is enhanced under 
the magnetic field parallel to the $x$-axis.

We discuss the physical reason
for the field enhancement of the AFM correlation:
First, the uniform magnetization induced by 
${\bf B}\parallel {\hat x}$
will reduce the AFM correlation
along the $x$-direction.
This leads to the enhancement of $\chi_{y}^{\rm max}$ 
by contraries,
as a result of solving the conflict between
spin-fluctuations with different components.
The increase in $\chi_{y}^{\rm max}$ will be 
more prominent in lower dimensional systems
because the reduction of $T_{\rm N}$ due to fluctuations
is large in general.
Note that the reduction of the staggered moment at $T=0$
owing to the quantum fluctuations
is approximately 40\%(15\%) in two (three) dimensional 
$S=1/2$ Heisenberg model without a magnetic field.

Consistently with the above discussion,
$\chi_{y(z)}^{\rm max}$ increases whereas 
$\chi_{x}^{\rm max}$ decreases under ${\bf B}\parallel {\hat x}$
in the present model by the FLEX approximation.
We have checked that this is a universal behavior
in two-dimensional systems close to the AFM-QCP,
by studying various types of Hubbard model.
Here, we briefly discuss the 
self-energy effect for susceptibilities:
When ${\bf B}=0$,
$\chi_{\uparrow,\downarrow}^0(\q,0)$ by the FLEX approximation
is reduced from the RPA's value
because of the reduction of the DOS, which is 
caused by the large Im$\Sigma$ under strong spin-fluctuations.
Considering that
$\Sigma(k) \approx U^2T\sum_{q}\sum_\a^{x,y,z} 
 \chi_{\a}(q) G^0(k+q)$,
the change in $\chi_{\uparrow,\downarrow}^0(\q,0)$
within the lowest order with respect to the self-energy
is given by
\begin{eqnarray}
\delta'\chi_{\uparrow,\downarrow}^0(\q,0)
&\approx& -T^2\sum_{k,q'} G^0(k)^2 G^0(k+\q) G^0(k+q')
 \nonumber \\
& &\times 2U^2 (\chi_{x}(q') + 2\chi_{y}(q'))
 \label{eqn:dc1}.
\end{eqnarray}
Because it is negative,
$\chi_{y}^{\rm max}$ in the FLEX approximation
becomes smaller than that in RPA.
Once ${\bf B}\parallel {\hat x}$ is applied,
the reduction in $\chi_{x}(q')$ 
owing to the field-induced uniform magnetization
will make
$|\delta'\chi_{\uparrow,\downarrow}^0(\Q,0)|$ smaller.
Consequently,
$\chi_y^{\rm max}$ increases in proportion to $B^2$
as long as only the self-energy effect is taken into account.


On the other hand,
unphysical results are obtained 
by RPA, where
all $G$'s in Eqs. (\ref{eqn:chi-xx})-(\ref{eqn:Sigma})
are replaced with $G_0$'s.
In the case $n=0.90$,
both $\chi_{x}^{\rm max}$ and $\chi_{y}^{\rm max}$
by RPA increase with $B$ as shown in Fig. \ref{fig:chi-FLEX},
possibly reflecting the fact that
the FS is close to the van-Hove singularity.
On the contrary,
both $\chi_{x}^{\rm max}$ and $\chi_{y}^{\rm max}$ 
decreases with ${\bf B}$ when $n=1.20$.
Thus, results given by the RPA are not universal, 
depending sensitively on the shape of the FS.
As a result,
the self-energy effect 
included in the FLEX approximation is indispensable 
in reproducing the physically reasonable behavior of the two-dimensional 
nearly AFM metals (i.e., $\alpha_{\rm S}\simge 0.98$)
under a magnetic field.

\begin{figure}
\begin{center}
\epsfig{file=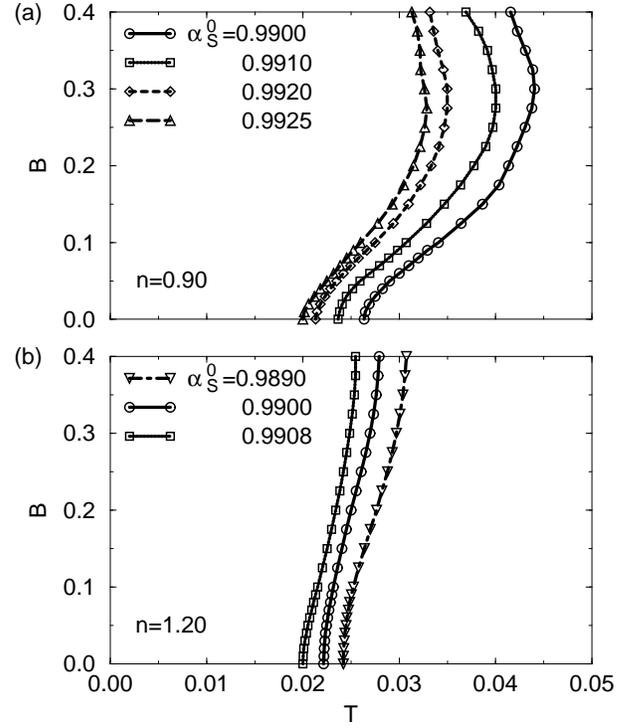,width=8cm}
\end{center}
\caption{
Obtained phase diagram for $T_{\rm N}$ vs $B$
for various $\a_{\rm S}^0$'s.
}
  \label{fig:phase}
\end{figure}

In the next stage,
we study the magnetic-field dependence of 
the N{\'e}el temperature $T_{\rm N}$
by assuming a weak three-dimensional coupling
 \cite{Kino-Kontani,Kino-Kontani2}.
To simplify the analysis,
we define $T_{\rm N}$ 
in the presence of the magnetic field
under the condition that
$\max_{\q}U\chi_{\uparrow,\downarrow}^0(\q,0)=\a_{\rm S}^0$,
where $\a_{\rm S}^0$ is a constant which 
is slightly smaller than unity.
By putting $\a_{\rm S}^0 =1-J_\perp/U \sim 0.99$
($J_\perp$ denotes the interlayer magnetic coupling strength),
we obtained reasonable N{\'e}el temperatures
of $\kappa$-(BEDT-TTF)$_2$X and TMTSF
based on the dimer model
 \cite{Kino-Kontani,Kino-Kontani2}.
Figure \ref{fig:phase}
shows the field dependence of $T_{\rm N}$ 
given by the FLEX approximation,
for several $\a_{\rm S}^0$'s.
We find that the 
field enhancement of the N{\'e}el temperature occurs
in nearly AFM metals in two dimensions,
which has been pointed out in the present work
for the first time.
In Fig. \ref{fig:phase}, $T_{\rm N}$ starts to increase
in proportion to $B^2$,
and it almost saturates at approximately $B^\ast\sim0.3$ when $n=0.90$.
This result also means that
the system approaches the AFM-QCP
by applying a magnetic field.
The increment in $T_{\rm N}$ is larger when $n=0.90$,
reflecting the closeness to the van-Hove singularity.

Here, it is notable that
in the antiferromagnetic isotropic Heisenberg chain 
under the magnetic field along the $x$-axis,
$\langle S_i^x S_j^x \rangle -M^2 \propto 
 (-1)^{i-j}|i-j|^{-1/\eta} \cos(2\pi M(i-j))$
and $\langle S_i^y S_j^y \rangle 
 \propto (-1)^{i-j}|i-j|^{-\eta}$,
where $\eta$ decreases from unity with the magnetic field
 \cite{Heisenberg}.
Their field dependencies are consistent with the
present study of a two-dimensional Hubbard model.
In the XXZ-Heisenberg chain,
an infinitely small magnetic field along the $x$-axis induces
the staggered magnetization of the $y$-component 
in the case of $J_z<J_x$
 \cite{XXZ-Caux,XXZ-Dmitriev}.
In the opposite case, $J_z>J_x$,
the staggered magnetization along the $z$-axis,
which exists without the field,
is enhanced by ${\bf B}\parallel {\hat x}$
 \cite{XXZ-Hieida}.
We also point out that Fukusima and Kuramoto
studied a localized electron model
with interactions between quadrupole moments
by a local approximation,
and found the field enhancement in $T_{\rm Q}$
due to the suppression of fluctuations
 \cite{Kuramoto2}.

Note that the field-induced SDW is
realized in the quasi-one dimensional metal, TMTSF,
owing to the orbital motion of electrons, 
free from the Zeeman effect
 \cite{TMTSF}.
However, various characteristics of the 
FI-AFM in CeRhIn$_5$ 
do not coincide with that observed in TMTSF.
In fact, CeRhIn$_5$ possesses both cylindrical and
spherical FS's.
They are naturally explained in terms of 
the Zeeman effect as discussed in the present study.

We now discuss the experimental results of
CeMIn$_5$ in the present study.
The bandwidth of the present model is $\sim 10$.
If we estimate the renormalized quasiparticle bandwidth
of CeMIn$_5$ to be $\sim$1000 K
 \cite{Yamada-rev-com},
the temperature $T=0.02$ corresponds to $\sim$2 K,
which is close to $T_{\rm c}$ in CeCoIn$_5$
 \cite{Tc}.
The magnetic field $B=0.1$ in the present work
corresponds to $\sim 5$ T
for the $M=\pm 5/2$ Kramers doublet (KD),
because the Zeeman energy for Ce$^{3+}$
is $(6/7)\mu_{\rm B}M H$
 \ ($6/7$ is the $g$-value of Ce$^{3+}$). 
Note that the renormalization factor
averaged over the FS is $0.217$
in the present FLEX approximation 
for $U=5$ at $T=0.02$.
$T_{\rm N}$ in CeRhIn$_5$ continues to increase
with the magnetic field parallel to the $ab$-plane,
at least below 9 T; 
$T_{\rm N}=3.8$K at 0T, and 
$T_{\rm N}(9 T)-T_{\rm N}(0 T) \approx 0.15$ K
 \cite{TN1,TN2}.
Whereas $T_{\rm N}$ decreases monotonically
when ${\bf B}\parallel {\hat c}$, as is observed
in usual 3D heavy Fermion systems.
This is naturally understood because 
the orbital motion of electrons, 
which is absent in the present study
where ${\bf B}$ is parallel to the 2D system,
will destroy the AFM state to obtain the 
energy due to the Landau diamagnetism.

Furthermore, 
we discuss the anisotropy of ${\hat \chi}(q)$ in CeRhIn$_5$:
The lowest KD of Ce$^{3+}$-ion in CeRhIn$_5$ is $\Gamma_7^{(2)}$;
$|z;\pm\rangle \equiv \beta|M_z\! = \!\pm 5/2\rangle 
 - \a |M_z \! =\! \mp 3/2\rangle$
 \cite{neutron1,neutron2},
which is approximately 70 K lower than the second lowest KD.
If we put $(\a,\beta)\approx(0.44,0.9)$
 \cite{neutron2},
$\langle z;\pm|J_{z}|z;\pm \rangle = \pm(2.5\beta^2-1.5\a^2)
 \approx \pm 1.74$.
On the other hand,
$\langle x;\pm|J_{x}|x;\pm \rangle = \pm\sqrt{5}\a\beta
 \approx \pm 0.885$,
where
$|x;\pm \rangle \equiv (|z;+\rangle \mp |z;-\rangle)/\sqrt{2}$.
Then, the anisotropy of the susceptibility of a single Ce$^{3+}$-ion
is $\chi_a/\chi_c \approx 1.74/0.885 =1.97$,
which is similar to the experimental ratio. 
On the other hand,
several neutron experiments on CeRhIn$_5$ revealed that
the magnetic moments on Ce sites lie
on the $ab$-plane below $T_{\rm N}$,
whose effective moment is $\mu_{\rm eff}=0.264\mu_{\rm B}$
 \cite{neutron1,neutron2}. 
This suggests that
the antiferromagnetic RKKY interaction between 
nearest neighbor Ce sites is XY-like; $J_{a,b}> J_c$ 
 \cite{Shiba}.
Then, the magnetic field along the $a$-axis
will enhance the AFM correlation along the $b$-axis
as a result of the reduction of fluctuations,
which is similar to the behavior of the XXZ-Heisenberg chain 
under ${\bf B}$ 
 \cite{XXZ-Caux,XXZ-Dmitriev}.
In fact, $\mu_{\rm eff}$ is much smaller than 
$(6/7)\mu_{\rm B}\langle x;+|J_{x}|x;+ \rangle 
 \approx 0.76\mu_{\rm B}$,
which suggests that the quantum fluctuations are strong
in CeRhIn$_5$, reflecting its two-dimensionality.
As a result, the field-enhancement in $T_{\rm N}$
observed in CeRhIn$_5$ is well understood
in terms of
the reduction of spin-fluctuations by a magnetic field.
It is a future research problem to take the Kondo effect 
into account beyond the FLEX approximation.

In summary,
on the basis of the FLEX approximation,
we found the field-induced antiferromagnetism
in a two-dimensional Hubbard model,
as a result of solving the conflict
between fluctuation with different directions
by a magnetic field.
This phenomenon is expected to be prominent and universal
in the vicinity of the AFM-QCP in lower dimensional systems,
irrespective of the fact that
the field-induced uniform magnetization
tends to decrease the AFM moments.
The induced AFM moments are almost on the plane
perpendicular to the applied magnetic field,
to earn the Zeeman energy by canting.
Experimentally,
the field-induced increment in $T_{\rm N}$ 
will be more prominent when ${\bf B}$ is 
parallel to the 2D system, because the reduction of 
$T_{\rm N}$ caused by the orbital motion effect 
(Landau quantization) is absent.

As for two-dimensional organic metals, the field-induced transition
from the paramagnetic metal to the AF insulator 
is found in $\kappa$-(ET)$_2$Cu[N(CN)$_2$]Cl
under a pressure,
below $T_{\rm c0}=13$ K and above $H_{c2}$
 \cite{Kanoda}.
Also, field-induced SDW due to the Zeeman effect
is expected to be realized in $\tau$-phase organic metals
 \cite{Murata}.
These phenomena will be explained by the present mechanism
 \cite{future}.

Finally, we note that the
present results by the FLEX approximation
seems reasonable in terms of the 
fluctuation-dissipation theorem;
$\langle {S_\a}^2 \rangle = \sum_\q\int_0^\infty \frac{d\w}{\pi}
 {\rm Im}\chi_\a(q)\coth(\frac{\w}{2T})$ ($\a=x,y,z$), and 
$\sum_\a^{x,y,z}\langle {S_\a}^2 \rangle \approx \frac{3n}{4}$
in strongly correlated systems.
When $\chi_\a(Q)$ grows,
$\langle {S_\a}^2 \rangle$ will increase
(especially in 2D systems).
Thus, $\chi_{y(z)}(Q)$ will increase when 
$\chi_x(Q)$ decreases by ${\bf B}\parallel {\hat x}$
 \cite{future}.

We are grateful to K. Yamada, D.S. Hirashima,
K. Miyake, Y. Kuramoto, M. Tsuchiizu and Y. Matsuda 
for useful discussions.

\vspace{5mm}



\end{multicols}


\begin{thebibliography}{99}
\bibitem{Yamada-rev-com}
In the LDA study of CeMIn$_5$,
the bandwidth of heavy-quasiparticles
given by the $c$-$f$ mixing is larger than 1000K.
Here, we analize the ``renormalized band'' 
owing to the many-body effect,
using the FLEX approximation in terms of the
``residual interaction''; see
 ref.\cite{Yamada-rev}.
\bibitem{Tc}
By solving the Eliashberg equation within the FLEX approximation,
$T_{\rm c}\approx 0.02$ with $d_{x^2\mbox{-}y^2}$-wave symmetry
is obtained in the preset model.
\end{thebibliography}
\end{document}